\documentclass{article}
\usepackage[T1]{fontenc}
\usepackage{tikz}
\usepackage{circuitikz}
\usetikzlibrary{circuits.logic.US}
\usepackage{graphicx}
\usepackage{nicefrac}
\usepackage{url}

\begin{document}
\title{\bf Pseudo-random generators using linear feedback shift registers with output extraction}
\author{Holger~Nobach\\[1.5mm]{}\small Max Planck Institute for Dynamics and Self-Organization\\{}\small 
Am Fa\ss{}berg 17, 37077 G\"ottingen, Germany\\[1.5mm]{}\small holger.nobach@nambis.de}%
\date{\today}

\maketitle

\begin{abstract}
The use of three extractors, fed by linear feedback shift registers (LFSR)
for generating pseudo-random bit streams is investigated. 
Specifically, a standard LFSR is combined with a von Neumann extractor, 
a modified LFSR, extended by the all-zero state, is combined with an output logic,
which translates every three bits from the LFSR into up to two output bits and a run 
extraction of the input bit stream into single output bits are investigated. The latter two 
achieve better efficiency in using bits from the primary bit stream, the last one reaches 50\%.
Compared to other generator logics, the three extractors investigated
are less performant in terms of their cryptographic strength. 
However, the focus of this report is on the quality of the pseudo-random bit stream in comparison to 
really random bits and on the efficiency of using the bits of the primary stream from the LFSR
and generating valid output bits, while fulfilling a minimum cryptographic strength only, 
beyond that of the pure LFSR.
\end{abstract}

%Keywords:
%linear feedback shift register, pseudo-random generator, extractor,
%linear complexity, symmetric-key encryption, stream cipher 

\section{Introduction}
Linear feedback shift registers (LFSR) are a broad class of pseudo-random bit stream generators,
which are widely used because of their simplicity and their good signal characteristics in terms of periodicity
with long periods, correlation, run lengths and balance of zeros and ones. Unfortunately, they are 
cryptographically weak --- for an $n$-bit long shift register a sequence of $2n$ stream bits is 
sufficient to construct at least an adequate linear feedback circuit and the appropriate initial state 
of the register, which can reproduce the complete bit sequence \cite{massey_69}.

To improve the cryptographic strength of a pseudo-random stream generator based on LFSRs, 
three basic concepts are generally used: (non-linear) output filters, (non-linear) combinations 
of multiple LFSRs or clock controlled LFSRs. Generators based on non-linear feedback shift registers 
are the ultimate solution, however, this is beyond the present discussion.
Non-linear output filters and non-linear combinations of LFSRs often lead to biased bit streams 
with unequal distributions of zeros and ones.
To obtain an unbiased output bit stream from a biased input, the von Neumann extractor \cite{vonneumann_51} 
can be used, which translates the four possible combinations of two consecutive input bits
into three possible states of the output one, zero or invalid/discarded, where the valid output bits
are obtained from combinations with identical probabilities, finally leading to an unbiased output sequence.

Alternatively, a self-shrinking generator \cite{meier_staffelbach_94} also introduces cryptographic strength
to an unbiased but weak bit sequence from an LFSR. The self-shrinking logic can be interpreted as a
certain form of a clock controlled LFSR, where some bits from the primary stream are discarded
by pulsing the LFSR multiple times between the bits selected and given out.
The von Neumann extractor and the self-shrinking generator, both read two consecutive bits from 
the primary bit stream at once and generate either one valid output bit, zero or one, or no valid output bit.
The difference between these two extraction logics is the mapping of the four input combinations 
to the output states. However, the von Neumann extractor seems a probable alternative to the 
self-shrinking logic to obtain cryptographic strength, even if the feature of the von Neumann extractor 
to correct a biased bit stream is not needed in combination with LFSRs.

Since the von Neumann extractor generates one output bit from two input bits, and this on average 
in half of the bit combinations, while the other half of the bit combinations gets discarded, 
the efficiency of such a generator in generated bits per primary input bit is approximately $\nicefrac{1}{4}$.
Options to increase the efficiency of generating output bits from input bits and lowering the number of discarded bits
are to increase the size of input chunks and generate more than one bit out of them \cite{elias_72}
or to reuse discarded bits from the input stream \cite{peres_92} to produce more output bits from a given input sequence.

In this report, an LFSR was first combined with a von Neumann extractor. Some basic statistical investigations are given.
The ideas to increase the chunk size and to produce multiple bits and that to reuse bits from the input stream 
are also being pursued, however in a different way than in the publications above.
Since the input stream coming from the LFSR is already unbiased,
the extractors do not have to worry about eliminating any bias. So the focus of the extraction logic defined
by the mapping of input onto output bit combinations can be on the quality of the generated bit stream 
compared to that of real random noise at a minimum cryptographic strength, at least above that of the pure LFSR.
Particularly, an extraction scheme is proposed, generating up to two output bits from every three input
bits coming from the feeding LFSR and reaching an efficiency of $\nicefrac{5}{12}$.
Furthermore, an extractor is investigated, deriving one valid bit or an invalid one like the von Neumann extractor
from three input bits, however, the LFSR is pulsed only once, reading one new bit from the LFSR and recycling 
the previous two bits, reaching an efficiency of $\nicefrac{1}{2}$.

All these logics derive their cryptographic strength from the fact that the output stream withholds  
information on how many bits of the primary stream have been discarded or have not been used in between two consecutive 
bits of the output stream and which values they have had. On the other side, this means that the more efficient the  
logic uses the input bits, the weaker the entire generator is. However, the focus of the present investigation 
is more on the efficiency of the generators having at least some cryptographic strength beyond the pure LFSR. 
For high cryptographic demands, other generators will be better suited \cite{deepthi_etal_09,masoodi_etal_12}.

Note that the term ``extractor'' is appropriate for the von Neumann and for the run extractors only, 
since they extract bits from the primary LFSR stream and discards others. The three-bit logic uses a translation table
instead, which does not correspond directly with the bits from the primary stream. However, for readability reason,
the term ``extractor'' is generally used for all three logics.

The Python programs used for this test 
with the three generator logics are openly available at \url{http://www.nambis.de/publications/arxiv24a.html}.

\section{Standard Linear Feedback Shift Register}
\label{sec:sLFSR}

\begin{figure}
\begin{center}    % -> \centerline
\begin{tikzpicture}[circuit logic US, >=latex]
  \ctikzset{flipflops/scale=0.75}
  \matrix[column sep=2mm]
  {
     &&&&\node [xor gate, point left, scale=0.75] (a11) {}; && \node [xor gate, point left, scale=0.75,yshift = -1.025mm] (a13) {};
     & \node [xor gate, point left, scale=0.75,yshift = -2.05mm] (a14) {};\\[5mm]
     \node [flipflop D,scale=0.5] (r1) {1}; & \node [flipflop D,scale=0.5] (r2) {2}; & \node [flipflop D,scale=0.5] (r3) {3}; & 
     \node {...}; &
     \node [flipflop D,scale=0.5] (r11) {11}; & \node [flipflop D,scale=0.5] (r12) {12}; & \node [flipflop D,scale=0.5] (r13) {13}; &
     \node [flipflop D,scale=0.5] (r14) {14}; & \node [flipflop D,scale=0.5] (r15) {15}; & \node [flipflop D,scale=0.5] (r16) {16};\\
  };
  \draw (5.575,-0.25) node (aus) {$a_i$};
  \draw (-5.8,-1.25) node (ein) {CLK};

  \draw (r1.pin 6) -- ++(right:2mm) |- (r2.pin 1);
  \draw (r2.pin 6) -- ++(right:2mm) |- (r3.pin 1);
  \draw (r3.pin 6) -- ++(right:1.5mm);
  \draw (r11.pin 1) -- ++(left:1.5mm);
  \draw (r11.pin 6) -- ++(right:2mm) |- (r12.pin 1);
  \draw (r12.pin 6) -- ++(right:2mm) |- (r13.pin 1);
  \draw (r13.pin 6) -- ++(right:2mm) |- (r14.pin 1);
  \draw (r14.pin 6) -- ++(right:2mm) |- (r15.pin 1);
  \draw (r15.pin 6) -- ++(right:2mm) |- (r16.pin 1);
  \draw (r16.pin 6) -- ++(right:2mm) |- (aus);
  \draw (r16.pin 6) -- ++(right:1mm) |- (a14.input 2);
  \filldraw (r16.pin 6) ++(right:1mm) circle (0.35mm);
  \draw (r14.pin 6) [short,*-] -- ++(right:1mm) |- (a14.input 1);
  \filldraw (r14.pin 6) ++(right:1mm) circle (0.35mm);
  \draw (a14.output) -- ++(left:2mm) |- (a13.input 2);
  \draw (r13.pin 6) -- ++(right:1mm) |- (a13.input 1);
  \filldraw (r13.pin 6) ++(right:1mm) circle (0.35mm);
  \draw (a13.output) -- ++(left:2mm) |- (a11.input 2);
  \draw (r11.pin 6) -- ++(right:1mm) |- (a11.input 1);
  \filldraw (r11.pin 6) ++(right:1mm) circle (0.35mm);
  \draw (r1.pin 1) -- ++(left:1mm) |- (a11.output);
  \draw (r1.pin 3) -- ++(left:1mm) |- (ein);
  \filldraw (r1.pin 3) ++(left:1mm) ++(down:3.7mm) circle(0.35mm);  
  \draw (r2.pin 3) -- ++(left:1mm) |- (ein);
  \filldraw (r2.pin 3) ++(left:1mm) ++(down:3.7mm) circle(0.35mm);  
  \draw (r3.pin 3) -- ++(left:1mm) |- (ein);
  \filldraw (r3.pin 3) ++(left:1mm) ++(down:3.7mm) circle(0.35mm);  
  \draw (r11.pin 3) -- ++(left:1mm) |- (ein);
  \filldraw (r11.pin 3) ++(left:1mm) ++(down:3.7mm) circle(0.35mm);  
  \draw (r12.pin 3) -- ++(left:1mm) |- (ein);
  \filldraw (r12.pin 3) ++(left:1mm) ++(down:3.7mm) circle(0.35mm);  
  \draw (r13.pin 3) -- ++(left:1mm) |- (ein);
  \filldraw (r13.pin 3) ++(left:1mm) ++(down:3.7mm) circle(0.35mm);  
  \draw (r14.pin 3) -- ++(left:1mm) |- (ein);
  \filldraw (r14.pin 3) ++(left:1mm) ++(down:3.7mm) circle(0.35mm);  
  \draw (r15.pin 3) -- ++(left:1mm) |- (ein);
  \filldraw (r15.pin 3) ++(left:1mm) ++(down:3.7mm) circle(0.35mm);  
  \draw (r16.pin 3) -- ++(left:1mm) |- (ein);

\end{tikzpicture}
\end{center}
\caption{Standard Fibonacci-type 16-bit LFSR with a period of $2^{16}-1$ internal states}
\label{fig:sLFSR}
\end{figure}
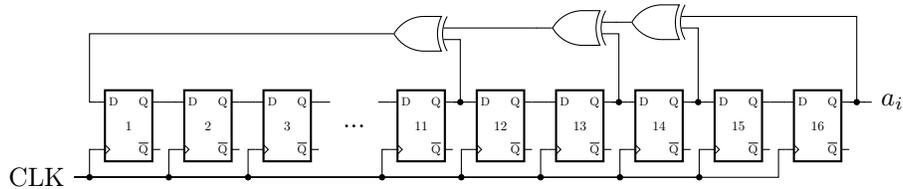

For the present investigation a 16-bit standard Fibonacci-type LFSR (sLFSR16, Fig.\ \ref{fig:sLFSR})
is used to generate the primary bit stream $a_i$.
A primitive polynomial for this LFSR \cite{ward_molteno_07} gives a possible structure to obtain the 
maximum cycle length of $2^{16}-1$ states by feeding back certain bits of the register to its input 
using three XOR gates.
For practical applications the length $n$ of the LFSR should be larger, yielding $2^n-1$ steps long periods of the primary bit stream 
and finally long periods of the output bit stream after the following extraction. So far, 160-bit LFSRs seem to be a
good choice, where the initialization of the LFSR can be done with an SHA-1 hash obtained from a secret password.

\section{Von Neumann extractor}
\label{sec:vonNeumann_generator}

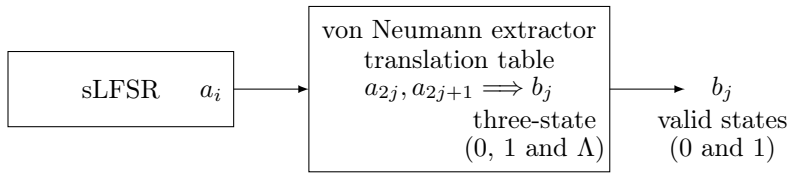
\begin{figure}
\begin{center}
\begin{tikzpicture}[>=latex]
\draw (-6,-0.5) rectangle (-3,0.5);
\draw (-4.5,0) node {sLFSR};
\draw (-3.3,-0.05) node {$a_i$};
\draw[->] (-3,0) -- (-2,0); 
\draw (-2,-1.1) rectangle (2,1.1) (0,0.8) node{von Neumann extractor} (0,0.4) node{translation table};
\draw (0,0) node{$a_{2j},a_{2j+1}\Longrightarrow b_j$};
\draw (1,-0.4) node {three-state};
\draw (1,-0.8) node {($0$, $1$ and $\Lambda$)};
\draw[->] (2,0) -- (3,0);
\draw (3.5,0) node {$b_j$};
\draw (3.5,-0.4) node {valid states};
\draw (3.5,-0.8) node {($0$ and $1$)};

\end{tikzpicture}
\end{center}
\caption{Pseudo-random generator using a standard LFSR in combination with a von Neumann extractor}
\label{fig:vonNeumann_generator}
\end{figure}

To break the vulnerability of LFSRs to their predictability, shrinking the bit stream by selecting parts of the bits 
from the primary stream and discarding the others has been found to be an appropriate means.
The von Neumann extractor originally selects individual bits from a biased input bit stream to 
obtain an unbiased output. However, this extractor is also useful to introduce some cryptographic strength
to a bit stream obtained from an LFSR (Fig.\ \ref{fig:vonNeumann_generator}).
The translation table of the von Neumann extractor (Tab.\ \ref{tab:vonNeumann_generator}) assigns the three possible 
output states zero ($0$), one ($1$) and invalid ($\Lambda$) of the output bit $b_j$ to the two consecutive primary input bits 
$a_{2j}$ and $a_{2j+1}$.

Note that the shift register is pulsed twice for reading the two primary bits into the von Neumann extractor at once
to either generate one valid output bit or an invalid one. In any case, both primary input bits were consumed.
If the von Neumann extractor is recycling one bit of the input stream for the next output by shifting the bits of the LFSR 
only once per extractor output step, the output sequence will end in a regular pattern of alternating zeros and ones. 
Therefore, the von Neumann extractor always reads two bits from the LFSR at once and one full period of the shrunken output stream 
is reached after two full periods of the primary stream, because a standard LFSR has a period of the generated primary bit stream, 
which is not divisible by two.

\begin{table}
\caption{Translation table of the von Neumann extractor\label{tab:vonNeumann_generator}}
\begin{center}
\begin{tabular}{c|c}
\hline
\raisebox{-0.5mm}{input bits $a_{2j}$ and $a_{2j+1}$}&\raisebox{-0.5mm}{output bit $b_j$}\\[0.75mm]
\hline
\vbox{\vspace{0.5mm}\hbox{$\begin{array}{cc}0&0\\0&1\\1&0\\1&1\end{array}$}}&\vbox{\vspace{0.5mm}\hbox{$\begin{array}{c}\Lambda\\1\\0\\{}\Lambda\end{array}$}}\\
\hline
\end{tabular}
\end{center}
\end{table}

\section{Extended Linear Feedback Shift Register}
\label{sec:eLFSR}

\begin{figure}
\begin{center}
\begin{tikzpicture}[circuit logic US, >=latex]
  \ctikzset{flipflops/scale=0.75}

  \draw (-5.75,-0.3) node [xor gate, point down, scale=0.75] (a0) {};
  \draw (-4.8,0.5) node [nor gate, point left, logic gate inputs=nnnnnnnnnn, scale=0.5] (no) {};
  \matrix[column sep=2mm]
  {
     &&&&\node [xor gate, point left, scale=0.75] (a11) {}; && \node [xor gate, point left, scale=0.75,yshift = -1.025mm] (a13) {};
     & \node [xor gate, point left, scale=0.75, yshift = -2.05mm] (a14) {};\\[20mm]
     \node [flipflop D,scale=0.5] (r1) {1}; & \node [flipflop D,scale=0.5] (r2) {2}; & \node [flipflop D,scale=0.5] (r3) {3}; & 
     \node {\ldots}; &
     \node [flipflop D,scale=0.5] (r11) {11}; & \node [flipflop D,scale=0.5] (r12) {12}; & \node [flipflop D,scale=0.5] (r13) {13}; &
     \node [flipflop D,scale=0.5] (r14) {14}; & \node [flipflop D,scale=0.5] (r15) {15}; & \node [flipflop D,scale=0.5] (r16) {16};\\
  };
  \draw (5.675,-1) node (aus) {$a_i$};
  \draw (-5.9,-2) node (ein) {CLK};
  \draw (-1.575,-0.25) node {\ldots};
  \draw (-3.8,0.425) node {\ldots};

  \draw (r1.pin 6) -- ++(right:2mm) |- (r2.pin 1);
  \draw (r2.pin 6) -- ++(right:2mm) |- (r3.pin 1);
  \draw (r11.pin 6) -- ++(right:2mm) |- (r12.pin 1);
  \draw (r3.pin 6) -- ++(right:3mm);
  \draw (r11.pin 1) -- ++(left:1.5mm);
  \draw (r12.pin 6) -- ++(right:2mm) |- (r13.pin 1);
  \draw (r13.pin 6) -- ++(right:2mm) |- (r14.pin 1);
  \draw (r14.pin 6) -- ++(right:2mm) |- (r15.pin 1);
  \draw (r15.pin 6) -- ++(right:2mm) |- (r16.pin 1);
  \draw (r16.pin 6) -- ++(right:2mm) |- (aus);
  \draw (r16.pin 6) -- ++(right:1mm) |- (a14.input 2);
  \filldraw (r16.pin 6) ++(right:1mm) circle (0.35mm);
  \draw (r14.pin 6) [short,*-] -- ++(right:1mm) |- (a14.input 1);
  \filldraw (r14.pin 6) ++(right:1mm) circle (0.35mm);
  \draw (a14.output) -- ++(left:2mm) |- (a13.input 2);
  \draw (r13.pin 6) -- ++(right:1mm) |- (a13.input 1);
  \filldraw (r13.pin 6) ++(right:1mm) circle (0.35mm);
  \draw (a13.output) -- ++(left:2mm) |- (a11.input 2);
  \draw (r11.pin 6) -- ++(right:1mm) |- (a11.input 1);
  \filldraw (r11.pin 6) ++(right:1mm) circle (0.35mm);
  \draw (r1.pin 1) -- ++(left:1mm) -| (a0.output);
  \draw (a0.input 2) -- ++(up:1mm) |- (a11.output);
  
  \draw (r1.pin 3) -- ++(left:1mm) |- (ein);
  \filldraw (r1.pin 3) ++(left:1mm) ++(down:3.7mm) circle(0.35mm);  
  \draw (r2.pin 3) -- ++(left:1mm) |- (ein);
  \filldraw (r2.pin 3) ++(left:1mm) ++(down:3.7mm) circle(0.35mm);  
  \draw (r3.pin 3) -- ++(left:1mm) |- (ein);
  \filldraw (r3.pin 3) ++(left:1mm) ++(down:3.7mm) circle(0.35mm);  
  \draw (r11.pin 3) -- ++(left:1mm) |- (ein);
  \filldraw (r11.pin 3) ++(left:1mm) ++(down:3.7mm) circle(0.35mm);  
  \draw (r12.pin 3) -- ++(left:1mm) |- (ein);
  \filldraw (r12.pin 3) ++(left:1mm) ++(down:3.7mm) circle(0.35mm);  
  \draw (r13.pin 3) -- ++(left:1mm) |- (ein);
  \filldraw (r13.pin 3) ++(left:1mm) ++(down:3.7mm) circle(0.35mm);  
  \draw (r14.pin 3) -- ++(left:1mm) |- (ein);
  \filldraw (r14.pin 3) ++(left:1mm) ++(down:3.7mm) circle(0.35mm);  
  \draw (r15.pin 3) -- ++(left:1mm) |- (ein);
  \filldraw (r15.pin 3) ++(left:1mm) ++(down:3.7mm) circle(0.35mm);  
  \draw (r16.pin 3) -- ++(left:1mm) |- (ein);

  \draw (no.output) -- ++(left:1mm) -| (a0.input 1);

  \draw (r1.pin 6) -- ++(right:1mm) |- (no.input 1);
  \filldraw (r1.pin 6) ++(right:1mm) circle (0.35mm);
  \draw (r2.pin 6) -- ++(right:1mm) |- (no.input 2);
  \filldraw (r2.pin 6) ++(right:1mm) circle (0.35mm);
  \draw (r3.pin 6) -- ++(right:1mm) |- (no.input 3);
  \filldraw (r3.pin 6) ++(right:1mm) circle (0.35mm);

  \draw (r11.pin 6) -- ++(right:1mm) |- (no.input 6);
  \filldraw (r11.pin 6) ++(right:1mm) ++ (up:15.35mm) circle (0.35mm);
  \draw (r12.pin 6) -- ++(right:1mm) |- (no.input 7);
  \filldraw (r12.pin 6) ++(right:1mm) circle (0.35mm);
  \draw (r13.pin 6) -- ++(right:1mm) |- (no.input 8);
  \filldraw (r13.pin 6) ++(right:1mm) ++ (up:16.85mm) circle (0.35mm);
  \draw (r14.pin 6) -- ++(right:1mm) |- (no.input 9);
  \filldraw (r14.pin 6) ++(right:1mm) ++ (up:17.6mm) circle (0.35mm);
  \draw (r15.pin 6) -- ++(right:1mm) |- (no.input 10);
  \filldraw (r15.pin 6) ++(right:1mm) circle (0.35mm);

\end{tikzpicture}
\end{center}
\caption{Extended Fibonacci-type 16-bit LFSR with a period of $2^{16}$ internal states}
\label{fig:eLFSR}
\end{figure}

To obtain long periods of the output stream, the cycle length of the LFSR should be chosen to be coprime with the number of
input bits read by the following extractor. The von Neumann extractor reads two consecutive bits from the primary bit stream, 
and the cycle length of standard LFSRs of $2^n-1$ is coprime for any register length $n$.

In combination with the following three-bit extractor, this coprime requirement is not met for all lengths $n$ of standard LFSRs. 
To achieve maximum flexibility in choosing the length of the LFSR, the feedback input of the LFSR is slightly modified (Fig.\ \ref{fig:eLFSR}). 
By adding one extra XOR gate and a NAND gate with $n-1$ inputs, the period of the LFSR is extended by the otherwise missing state
of the standard LFSR, where all bits of the register are zero. The NAND gate together with the additional XOR gate 
at the input of the register inverts the input line if the first $n-1$ bits are zero. In this way, the all-zero state is inserted
into the sequence between the last standard state of the LFSR, with the last flip-flop one and all others zero, and the first 
standard state with the first flip-flop one and all others zero.

The obtained period of the bit stream from the extended LFSR (eLFSR) with $n$ registers is then $2^n$, which 
is always coprime with the three bits read by the following three-bit extractor.

\section{Three-bit extractor}
\label{sec:threebit_generator}

\begin{figure}
\begin{center}
\begin{tikzpicture}[>=latex]
\draw (-6.5,-0.5) rectangle (-3.5,0.5);
\draw (-5,0) node {eLFSR};
\draw (-3.8,-0.05) node {$a_i$};
\draw[->] (-3.5,0) -- (-2.5,0); 
\draw (-2.5,-1.1) rectangle (2.5,1.1) (0,0.8) node{three-bit extractor} (0,0.4) node{translation table};
\draw (0,0) node{$a_{3j},a_{3j+1},a_{3j+2}\Longrightarrow b_{2j},b_{2j+1}$};
\draw (1.5,-0.4) node {three-state};
\draw (1.5,-0.8) node {($0$, $1$ and $\Lambda$)};
\draw[->] (2.5,0) -- (3.5,0);
\draw (4.5,0) node {$b_{2j},b_{2j+1}$};
\draw (4.5,-0.4) node {valid states};
\draw (4.5,-0.8) node {($0$ and $1$)};

\end{tikzpicture}
\end{center}
\caption{Pseudo-random generator using an extended LFSR in combination with a three-bit extractor}
\label{fig:threebit_generator}
\end{figure}

\begin{table}
\caption{Translation table of the three-bit extractor\label{tab:threebit_generator}}
\begin{center}
\begin{tabular}{c|c}
\hline
\raisebox{-0.5mm}{input bits $a_{3j}$, $a_{3j+1}$ and $a_{3j+2}$}&\raisebox{-0.5mm}{output bits $b_{2j}$ and $b_{2j+1}$}\\[0.75mm]
\hline
\vbox{\vspace{0.5mm}\hbox{$\begin{array}{ccc}0&0&0\\0&0&1\\0&1&0\\0&1&1\\1&0&0\\1&0&1\\1&1&0\\1&1&1\end{array}$}}&
\vbox{\vspace{0.5mm}\hbox{$\begin{array}{cc}\Lambda&\Lambda\\0&\Lambda\\1&1\\0&1\\1&0\\1&\Lambda\\0&0\\{}\Lambda&\Lambda\end{array}$}}\\
\hline
\end{tabular}
\end{center}
\end{table}

In analogy to the von Neumann extractor and using the idea of reading larger chunks of bit to increase
the efficiency of the generator from \cite{elias_72}, a new translation table
(Fig.\ \ref{fig:threebit_generator} and Tab.\ \ref{tab:threebit_generator}) has been designed,
which uses every three bits of the primary LFSR bit stream to derive up to two valid output bits.
Two of the eight input combinations give no valid output bit, two others give one valid output bit 
and the last four input combinations give the four combinations with two valid output bits. 
The translation table does not consider bias correction, since the input stream from the LFSR 
is already unbiased. It is designed to maximize the variance of the output stream 
with respect to the variations of the input stream. Because this table is critical 
for the effort to break the pseudo-random generator, it could be subject to updates, 
if it is later found to be weak.

Since this extractor uses every three consecutive bits of the primary bit stream from the LFSR,
the extended LFSR (eLFSR) is used as input, yielding a period of the primary bit stream, which is coprime with the 
three bits read by the extractor.

\section{Run extractor}
\label{sec:run_generator}

\begin{figure}
\begin{center}
\begin{tikzpicture}[>=latex]
\draw (-6,-0.5) rectangle (-3,0.5);
\draw (-4.5,0) node {sLFSR};
\draw (-3.3,-0.05) node {$a_i$};
\draw[->] (-3,0) -- (-2,0); 
\draw (-2,-1.1) rectangle (2,1.1) (0,0.8) node{Run extractor} (0,0.4) node{translation table};
\draw (0,0) node{$a_j,a_{j+1},a_{j+2}\Longrightarrow b_j$};
\draw (1,-0.4) node {three-state};
\draw (1,-0.8) node {($0$, $1$ and $\Lambda$)};
\draw[->] (2,0) -- (3,0);
\draw (3.5,0) node {$b_j$};
\draw (3.5,-0.4) node {valid states};
\draw (3.5,-0.8) node {($0$ and $1$)};

\end{tikzpicture}
\end{center}
\caption{Pseudo-random generator using a standard LFSR in combination with a run extractor (three-bit interpretation)}
\label{fig:triplet_generator}
\end{figure}
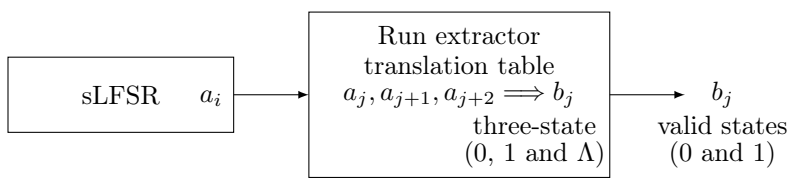

\begin{table}
\caption{Translation table of the run extractor (three-bit interpretation)\label{tab:triplet_generator}}
\begin{center}
\begin{tabular}{c|c}
\hline
\raisebox{-0.5mm}{input bits $a_j$, $a_{j+1}$ and $a_{j+2}$}&\raisebox{-0.5mm}{output bit $b_j$}\\[0.75mm]
\hline
\vbox{\vspace{0.5mm}\hbox{$\begin{array}{ccc}0&0&0\\0&0&1\\0&1&0\\0&1&1\\1&0&0\\1&0&1\\1&1&0\\1&1&1\end{array}$}}&
\vbox{\vspace{0.5mm}\hbox{$\begin{array}{c}\Lambda\\{}\Lambda\\0\\1\\0\\1\\{}\Lambda\\{}\Lambda\end{array}$}}\\
\hline
\end{tabular}
\end{center}
\end{table}

Since the von Neumann extractor reads two consecutive bits at once from the primary bit stream of the LFSR,
the efficiency of generating output bits out of the input stream is about $\nicefrac{1}{4}$ only. 
A higher efficiency would be the result, if input bits from the LFSR could be recycled by keeping 
one bit from the last reading step and read one new bit from the LFSR only for the next output.
Unfortunately, this results in a permanent sequence of alternating zeros and ones.
This behavior is originated in the fact that the von Neumann extractor uses the combination
of the two consecutive input bits to derive both, the selection decision and the data bit. 
The two bits must be different to generate a valid output bit. This is the case at the end of 
runs of either zeros or ones independent of their run lengths. Since the runs of zeros and ones 
occur alternately, they are finally shrunken down to an alternating series of single zeros and ones.

Using two of three consecutive bits from the LFSR to obtain the selection decision and using a third and separate bit
in the bit sequence as the data input, allows to generate one output bit (zero, one or invalid)
from each single clock impulse of the LFSR based on the three last output bits of the LFSR.
An appropriate sketch and its translation table are given in Fig.\ \ref{fig:triplet_generator}
and Tab.\ \ref{tab:triplet_generator}, where the third bit of each triplet is given out, 
if the two first bits are different, otherwise no bit is given out. This generator can also be interpreted 
as an output filter, which combines the three last flip-flops of the LFSR to one output bit (including the invalid state).
In half of the output combinations a valid bit is generated per clock impulse of the LFSR. Therefore, this
extractor logic achieves either approximately $\nicefrac{1}{2}$ efficiency for a combination with a standard LFSR
or exactly $\nicefrac{1}{2}$ for a combination with an extended LFSR. However, the output sequence is periodic within
one loop of the feeding LFSR, yielding exactly the same output sequence with both LFSR types.

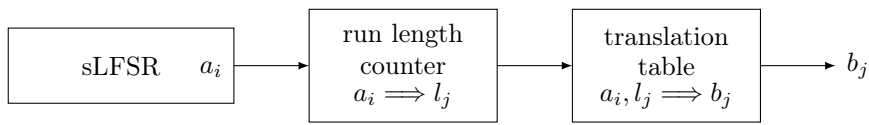
\begin{figure}
\begin{center}
\begin{tikzpicture}[>=latex]
\draw (-6.5,-0.5) rectangle (-3.5,0.5);
\draw (-5,0) node {sLFSR};
\draw (-3.8,-0.05) node {$a_i$};
\draw[->] (-3.5,0) -- (-2.5,0); 
\draw (-2.5,-0.75) rectangle (0,0.75) (-1.25,0.4) node{run length} (-1.25,0.0) node{counter};
\draw (-1.25,-0.4) node{$a_i\Longrightarrow l_j$};
\draw[->] (0,0) -- (1,0); 
\draw (1,-0.75) rectangle (3.5,0.75) (2.25,0.4) node{translation} (2.25,0.0) node{table};
\draw (2.25,-0.4) node{$a_i, l_j\Longrightarrow b_j$};
\draw[->] (3.5,0) -- (4.5,0);
\draw (4.8,0) node {$b_j$};
\end{tikzpicture}
\end{center}
\caption{Pseudo-random generator using a standard LFSR in combination with a run extractor (run length interpretation)}
\label{fig:runlength_generator}
\end{figure}

\begin{table}
\caption{Translation table of the run extractor (run length interpretation)\label{tab:runlength_generator}}
\begin{center}
\begin{tabular}{c|c|c}
\hline
\multicolumn{2}{c|}{\raisebox{-0.5mm}{run lengths $l_j$ of identical input bits $a_i\ldots a_{i+l_j-1}$}}&\raisebox{-0.5mm}{output bit $b_j$}\\[0.75mm]
\hline
\raisebox{-0.5mm}{\hbox to 3.5cm{\hfil $=1$\hfil}}&\raisebox{-0.5mm}{\hbox to 3.5cm{\hfil $0$\hfil}}&\raisebox{-0.5mm}{1}\\[0.5mm]
$\ge 2$&$0$&0\\
$=1$&$1$&0\\
$\ge 2$&$1$&1\\
\hline
\end{tabular}
\end{center}
\end{table}

Since the extractor logic discards the bits of the primary input stream as long as the two 
first of the three consecutive bits are equal, this logic could be interpreted alternatively as a run length dependent extractor.
A valid output bit can occur only at the end of a run, where the ending run gets broken after the last bit
by the first bit of the following run with the opposite value. This bit is denoted as the stop bit.
This new run of the new value can have either the run length one, then the next bit, following the
stop bit is opposite again. Or the new run starting with the stop bit has a
run length longer than one, then the next bit has the same value as the stop bit.

This way it is possible to re-interpret the extraction mechanism of the previous logic as a run length depending 
output extraction (Fig.\ \ref{fig:runlength_generator} and Tab.\ \ref{tab:runlength_generator}),
where the run lengths of zeros or ones generated by the LFSR are counted first.
A one is given out for a single zero bit or for a series of multiple ones from the LFSR and
a zero is given out for a single one bit or for a series of multiple zeros from the LFSR.
The difference between the two interpretations of the extraction logic is that the first one, using the three last bits 
from the LFSR generates the appropriate output bit at the beginning of the new run (more precisely at the 
second bit of the new run), while the run length interpretation reads the entire run and generates the 
appropriate output bit at the end of each run. Otherwise, the two generating mechanisms are equivalent, 
generating identical output streams, finally leading to the common designation ``run extractor''.

\section{Characterization}

For practical applications the length of the LFSRs should be large, resulting in long periods of the input streams and thus 
long periods of the output bit stream of the extractor and finally in a sufficient cryptographic strength 
of the pseudo-random generator. 
So far, 160-bit LFSRs seem to be a good choice for initialization, e.g.\ by an SHA-1 hash obtained from a secret password.
However, to characterize the performance of the generators, 16-bit versions of the LFSRs are used here, which has the advantage 
that a full period of the generator's output can be examined.
The following analysis characterizes the bit streams generated by three example generators, namely
\begin{enumerate}
\item[a)] a von Neumann extractor driven by every two consecutive bits from a standard 16-bit LFSR with a period of $2^{16}-1$ steps
(sLFSR16-vNe, see Figs.\ \ref{fig:sLFSR} and \ref{fig:vonNeumann_generator} in Secs.\ \ref{sec:sLFSR} and \ref{sec:vonNeumann_generator} 
and Tab.\ \ref{tab:vonNeumann_generator}),
\item[b)] a three-bit extractor driven by every three consecutive bits from an extended 16-bit LFSR with a period $2^{16}$ steps
(eLFSR16-3be, see Figs.\ \ref{fig:eLFSR} and \ref{fig:threebit_generator} in Secs.\ \ref{sec:eLFSR} and \ref{sec:threebit_generator}
and Tab.\ \ref{tab:threebit_generator}),
\item[c)] and a run extractor driven by single bits from a standard 16-bit LFSR with a period of $2^{16}-1$ steps
(sLFSR16-Re, see Figs.\ \ref{fig:sLFSR} and \ref{fig:triplet_generator} or \ref{fig:runlength_generator} 
in Secs.\ \ref{sec:sLFSR} and \ref{sec:run_generator} and Tab.\ \ref{tab:triplet_generator} resp.\ \ref{tab:runlength_generator}).
\end{enumerate}

%\subsection{Basic Counts}

The basic counts and statistics of the three example generators are given in Table~\ref{tab:basic_stats}.
The ones and zeros of the resulting output bit stream are perfectly balanced in all three cases. 
The sLFSR-vNe achieves $\nicefrac{1}{4}$ efficiency for long LFSRs (large number of registers $n$) 
and it is $\nicefrac{5}{12}$ for the eLFSR-3be. The sLFSR-Re achieves $\nicefrac{1}{2}$ efficiency 
for long LFSRs. Driving the run extractor by an extended LFSR yields an efficiency of exactly $\nicefrac{1}{2}$.

\begin{table}
\caption{Basic statistics of the sLFSR16-vNe, eLFSR16-3be and sLFSR16-RLbm\label{tab:basic_stats}}
\begin{center}
\begin{tabular}{l|c|c|c}
\hline
&\raisebox{-0.5mm}{sLFSR16-vNe}&\raisebox{-0.5mm}{eLFSR16-3be}&\raisebox{-0.5mm}{sLFSR16-Re}\\[0.5mm]
\hline
\raisebox{-0.5mm}{output period, counting \ldots}&&&\\[0.5mm]
\ \ all states (0, 1, $\Lambda$)&65535&131072&65535\\
\ \ valid states (0, 1)&32768&81920&32768\\
\ \ zeros&16384&40960&16384\\
\ \ ones&16384&40960&16384\\
efficiency (output bits&\vbox to 1mm{\hbox{$\frac{16384}{65535}\approx 25\%$}}&\vbox to 1mm{\hbox{$\frac{5}{12}\approx 42\%$}}&\vbox to 1mm{\hbox{$\frac{32768}{65535}\approx 50\%$}}\\
\ \ per input bit)&&&\\[0.25mm]
\hline
\end{tabular}
\end{center}
\end{table}

%\subsection{Basic Tests}

In \cite{FIPS140-1} statistical tests of randomness are defined that any random generator should pass. 
Table~\ref{tab:tests_tests} gives the values obtained and the requirements. The generators pass all the required tests.

\begin{table}
\caption{FIPS 140-1 statistical random number generator tests for randomness of the sLFSR16-vNe and eLFSR16-3be generators}
\label{tab:tests_tests}
\begin{center}
\begin{tabular}{l|r@{ }c@{ }l|c|c|c}
\hfil \vbox to 0mm{\hbox{test}}\hfil&\multicolumn{3}{c|}{\vbox to 0mm{\hbox{requirement}}}&sLFSR16&eLFSR16&sLFSR16\\
&&&&-vNe&-3be&-Re\\
\hline
\raisebox{-0.5mm}{monobit test}&\raisebox{-0.5mm}{$9654$}&\raisebox{-0.5mm}{$<X<$}&\raisebox{-0.5mm}{$10346$}&\raisebox{-0.5mm}{$10097$}&\raisebox{-0.5mm}{$9917$}&\raisebox{-0.5mm}{$10010$}\\
poker test&$1.03$&$<X<$&$57.4$&$25.958$&$9.613$&$12.954$\\
runs tests (zeros)&&&&&\\
\ \ run length 1&$2267$&$<X<$&$2733$ &$2539$&$2496$&$2536$\\
\ \ run length 2&$1079$&$<X<$&$1421$ &$1214$&$1240$&$1233$\\
\ \ run length 3&$502$&$<X<$&$748$   &$623$&$629$&$625$\\
\ \ run length 4&$223$&$<X<$&$402$   &$320$&$324$&$316$\\
\ \ run length 5&$90$&$<X<$&$223$    &$158$&$152$&$157$\\
\ \ run length 6+&$90$&$<X<$&$223$   &$147$&$166$&$153$\\
runs tests (ones)&&&&&\\
\ \ run length 1&$2267$&$<X<$&$2733$ &$2442$&$2515$&$2521$\\
\ \ run length 2&$1079$&$<X<$&$1421$ &$1296$&$1242$&$1244$\\
\ \ run length 3&$502$&$<X<$&$748$   &$626$&$656$&$629$\\
\ \ run length 4&$223$&$<X<$&$402$   &$313$&$295$&$327$\\
\ \ run length 5&$90$&$<X<$&$223$    &$166$&$152$&$140$\\
\ \ run length 6+&$90$&$<X<$&$223$   &$158$&$147$&$159$\\
long run test&\multicolumn{3}{c|}{$X=0$}&$0$&$0$&$0$\\[0.25mm]
\hline
&&&&\raisebox{-0.5mm}{all passed}&\raisebox{-0.5mm}{all passed}&\raisebox{-0.5mm}{all passed}
\end{tabular}
\end{center}
\end{table}

%\subsection{Output Patterns}

\begin{figure}
\centerline{\includegraphics{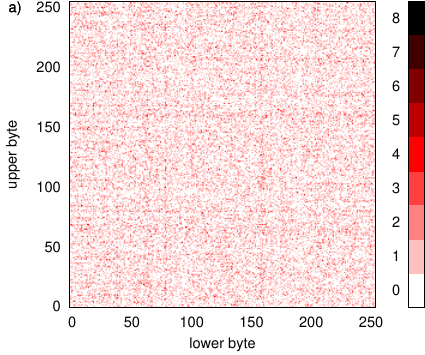}}\par
\centerline{\includegraphics{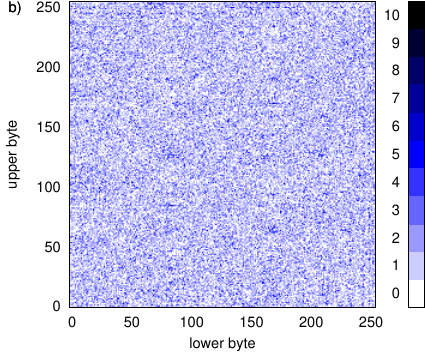}}\par
\centerline{\includegraphics{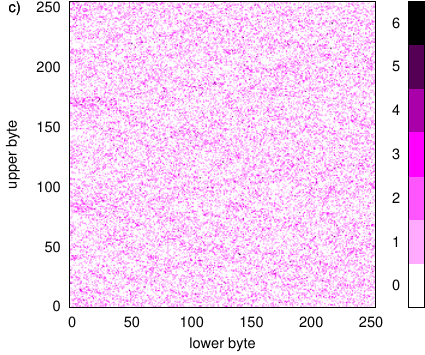}}
\caption{Counts of 16-bit patterns of a) the sLFSR16-vNe, b) the eLFSR16-3be and c) the sLFSR16-Re generators}
\label{fig:patterns_counts}
\end{figure}

\begin{figure}
\centerline{\includegraphics{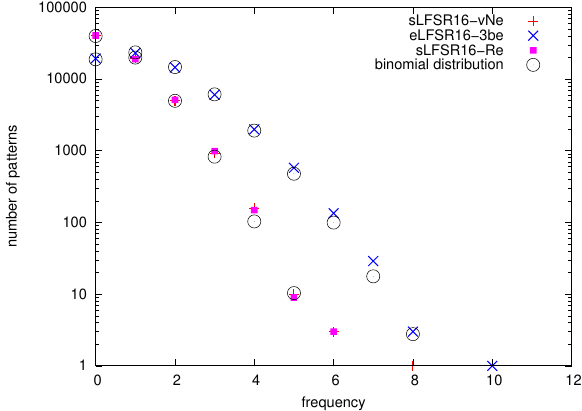}}
\caption{Histogram of 16-bit pattern repetitions of the sLFSR16-vNe, the eLFSR16-3be and the sLFSR16-Re generators 
in comparison to a binomial distribution}
\label{fig:patterns_hist}
\end{figure}

In contrast to the LFSR, where all possible patterns occur exactly once in one period of the primary bit stream,
after the extraction by the three example logics, some of the patterns occur multiple times and others 
do not occur in the output stream.
Fig.\ \ref{fig:patterns_counts} shows how many times the particular pattern occurs within one period of the output stream
for all possible 16-bit patterns, reaching from zero to eight, ten or six times respectively for some patterns for the three generators.
There is no indication of any deviation from a random occurrence of the patterns.

Fig.\ \ref{fig:patterns_hist} shows the histogram of the number of pattern repetitions,
namely how many of the possible 65536 patterns cannot be found in the output sequence, how many of them occur once,
how many occur twice, etc.. 
This can be seen as the orthogonal projection of the data in Fig.\ \ref{fig:patterns_counts}.
The empirical data have no significant deviation from the expected binomial distribution for random occurrences.

%\subsection{Run Lengths}

\begin{figure}
\centerline{\includegraphics{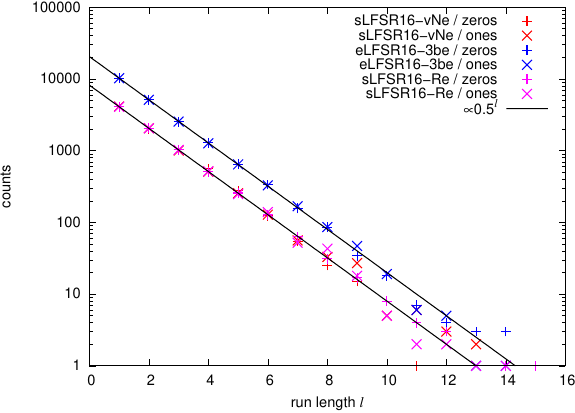}}
\caption{Run length histogram of sequences of zeros and ones for the sLFSR16-vNe, the eLFSR16-3be and the sLFSR16-Re generators
in comparison to the expected run lengths for random termination}
\label{fig:run_lengths}
\end{figure}

The observed run lengths of series of ones and zeros in Fig.\ \ref{fig:run_lengths} are well exponentially distributed.
For comparison the $0.5^l$ dependence of a perfectly random bit stream with $l$ the run length of series of ones or zeros
is also shown. 
Even if the total number of ones and zeros in one period are exactly balanced, their run lengths are differently distributed in detail,
but follow approximately the expected exponential decay.

%\subsection{Correlation}

\begin{figure}
\centerline{\includegraphics{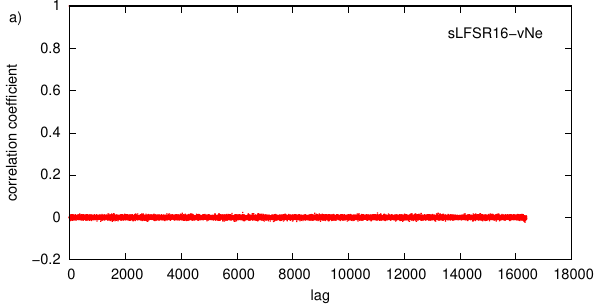}}\par
\centerline{\includegraphics{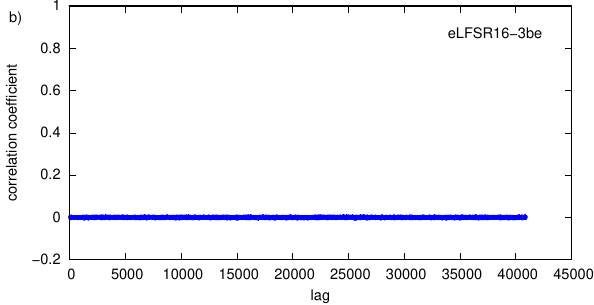}}\par
\centerline{\includegraphics{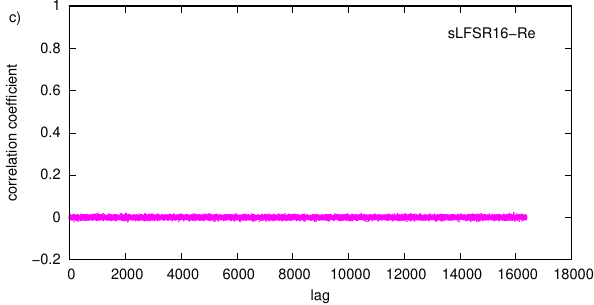}}
\caption{One-sided correlation functions of the output bit stream 
for a) the sLFSR16-vNe, b) the eLFSR16-3be and c) the sLFSR16-Re generators}
\label{fig:correlation}
\end{figure}

Since the output of the generators is periodic, the correlation function is periodic also.
Furthermore, the correlation function is symmetric. Therefore, Fig.\ \ref{fig:correlation} shows 
the correlation between zero lag and half of the period for the three generators.
Except for the correlation at lag zero, which must be one, the correlation for all other lags is around zero.
Further investigation shows that the standard deviation of the correlation values (except for the one value at lag zero) 
is $5.4\times 10^{-3}$ for the sLFSR-vNe, $3.5\times 10^{-3}$ for the eLFSR-3be and $5.5\times 10^{-3}$ for the sLFSR-Re. 
The expected value for a period of 32768 steps is $5.5\times 10^{-3}$ and that for a period of 81920 steps is $3.5\times 10^{-3}$, 
assuming perfect randomness of the output streams.

%\subsection{Power Spectra}

\begin{figure}
\centerline{\includegraphics{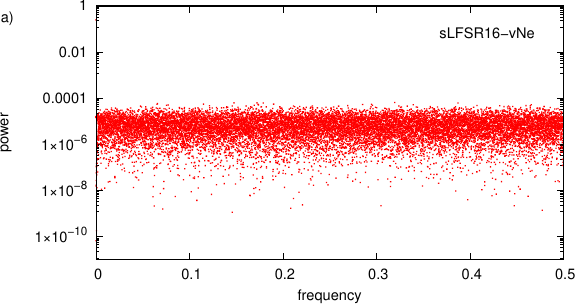}}\par
\centerline{\includegraphics{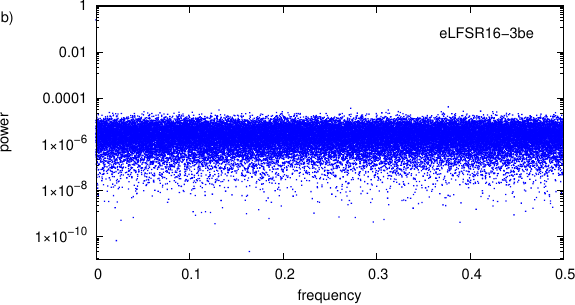}}\par
\centerline{\includegraphics{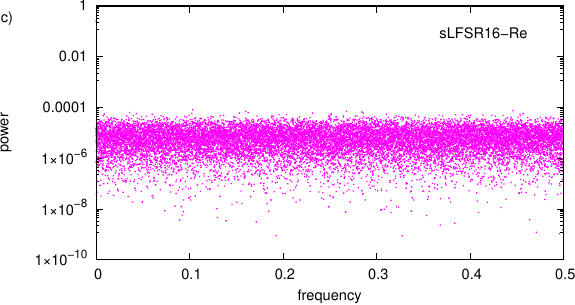}}
\caption{One-sided power spectra of the output bit stream 
for a) the sLFSR16-vNe, b) the eLFSR16-3be and c) the sLFSR16-Re generators}
\label{fig:powerspectra}
\end{figure}

In Fig.\ \ref{fig:powerspectra} the one-sided power spectra of the three generators are shown
in a logarithmic scale. The output stream has a mean value of $\nicefrac{1}{2}$. Therefore, the power at frequency zero
is $\nicefrac{1}{4}$. At all other frequencies the total noise power of $\nicefrac{1}{4}$ should  
be distributed over all frequencies leading to an expected power of $7.6\times 10^{-6}$ for
a period of 32768 steps and $3.1\times 10^{-6}$ for a period of 81920 steps.
The shown spectra of the investigated generators comply with a fully random bit stream.

%\subsection{Linear Complexity}

\begin{figure}
\centerline{\includegraphics{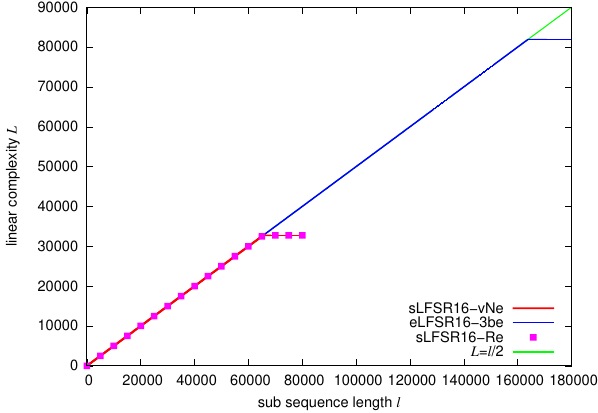}}
\caption{Linear complexity profiles of the sLFSR16-vNe, the eLFSR16-3be and the sLFSR16-Re generators
in comparison to an ideal random sequence generator}
\label{fig:linear_complexity}
\end{figure}

With the Berlekamp-Massey algorithm \cite{massey_69} the linear complexity $L$ of any subsequence 
of the generated bit streams can be derived as a function of the length $l$ of the 
subsequences, giving the linear complexity profile. Fig.~\ref{fig:linear_complexity} shows the 
linear complexity profiles for the three example generators, which all
follow the ideal $L=\nicefrac{l}{2}$ relation over a full period of their output streams.

%\subsection{Sequences of Discarded Bits}

\begin{figure}
\centerline{\includegraphics{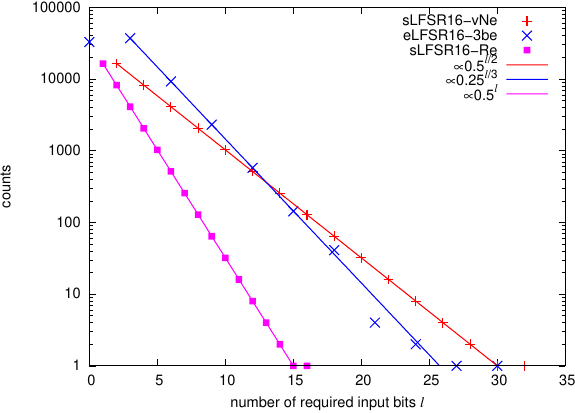}}
\caption{Histogram of number of required input bits to generate the next valid output bit 
for the sLFSR16-vNe, the eLFSR16-3be and the sLFSR16-Re generators 
in comparison to the expected lengths for random termination}
\label{fig:number_input}
\end{figure}

A disadvantage of the investigated generators in practical implementation with hardware circuits is that the number of input bits
required to generate the next output bit is not known in advance and changes randomly from one output bit to the next.
However, the varying number of primary bits from the LFSRs needed to produce one valid output bit,
on the other hand is an advantage in strengthening the bit stream in a cryptographic sense, because the output stream gives no
indication of how many bits of the input stream have been discarded between successive output bits.

Fig.~\ref{fig:number_input} shows the distribution of the number of primary bits required to generate 
the next output bit for the three generators investigated. 
Even if the mean values are known, the individual number of steps required is random with an exponential distribution,
which is also shown as a reference in the diagram.
While the sLFSR16-vNe requires at least two steps, the eLFSR16-3be generates two output bits at once in half 
of the input combinations. Therefore, there are also cases where no additional step of the eLFSR is required. 
However, this particular case, is outside the exponential law of the other values in the diagram.
The sLFSR16-Re generator reflects exactly the distribution of the run lengths of the LFSR and, 
therefore, perfectly follows the exponential decay until it reaches 16, which is the maximum 
possible run length corresponding to the length of the feeding LFSR.
The slopes of the distributions directly correspond to the efficiency of the generators given in Table~\ref{tab:basic_stats}.

\section{Conclusion}

The use of extractors or appropriate translation tables to obtain a shrunken bit stream from a primary bit stream of an LFSR
combines good statistical properties of the generated pseudo-random bit stream, such as a balanced distribution of ones and zeros, 
random run lengths and low correlation with a large linear complexity. 
Therefore, such combinations of an LFSR with extraction logics have potential as pseudo-random generators for cryptographic key streams.
The disadvantage of such an extractor is that it produces a shorter output stream with a significant number 
of discarded bits. At the same time, this is an advantage for its cryptographic strength, as it makes it difficult to reconstruct 
the primary stream from the output data.

The von Neumann extractor achieves about $25\%$ efficiency, meaning it generates one output bit from 
four bits generated by the LFSR on average. To maximize the period of the output stream, the von Neumann 
extractor is fed by a standard $n$ bit long LFSR with $2^n-1$ internal states.

With a similar concept, the three-bit extractor shown here, also has excellent statistical properties and has a
higher efficiency of nearly $42\%$ in generating the output stream, while again the cryptographic strength 
must be obtained by an unpredictable number of input bits discarded between two successive output bits
and a minimum coherence between the input and the output stream by an appropriate translation table.
With an appropriate translation table, a specific output pattern can be achieved from different inputs, which 
ideally have different lengths of the primary bit stream and a high variation in their input bit sequences.
Since the translation table is critical to the effort of breaking the pseudo-random generator, it could be subject to 
updates, if it is later found to be weak.
To obtain maximum periods of the output stream, the three-bit extractor is fed by an extended $n$ bit long LFSR 
with $2^n$ internal states.

The concept can be further extended to a four-bit extractor, again applied to a standard LFSR. In this case 
the efficiency can be further increased to $\nicefrac{17}{32}\approx 53\%$. To avoid weakening the cryptographic
strength, an appropriate translation table is required, which should be designed with special care, which may be added 
to the present investigations later.

With $50\%$, the run extractor reaches the highest efficiency among the three generator logics 
investigated here, with otherwise perfect random statistical properties. On the contrary, a higher efficiency in 
using the primary input bits to form the output stream comes with a lower cryptographic strength. 
Since each bit of the output stream is obtained from a run of either zeros or ones, the run extractor
is significantly weaker than the other two extraction logics investigated. Only the lengths of the respective runs 
are unknown. Since short runs are more probable than long runs, attacks could first try combinations of short runs
and successively test longer runs, which then are less probable and have less possible bit variations also.
Even if this extractor fulfills a certain minimum of cryptographic strength, it is not suitable 
for applications with serious demands in its cryptographic strength.
However, the present investigation had its focus more on the efficiency of the respective extractors 
and the quality of the generated pseudo-random bit streams than on their cryptographic strengths. 

The Python programs used for this test 
with the three generator logics are openly available at \url{http://www.nambis.de/publications/arxiv24a.html}.

\end{document}